\documentclass[useAMS, usenatbib]{mn2e}
 
\title[Self-regulated Black Hole Accretion]{Self-regulated black hole accretion, the $M-\sigma$ relation, and the growth of bulges in galaxies}

\author[M. C. Begelman, B. B. Nath]
{Mitchell C. Begelman$^{1}$\thanks{E-mail: mitch@jila.colorado.edu (MB); biman@rri.res.in (BN)}\thanks{Also at Department of Astrophysical and Planetary Sciences, University of Colorado at Boulder.} and Biman B. Nath$^{1,2}$\footnotemark[1]\thanks{JILA Visiting Fellow.}\\
$^1$JILA, University of Colorado, Boulder, CO 80309-0440, USA \\ $^2$Raman Research Institute, Sadashivanagar, Bangalore
-560080, India  }

\begin{document}
\maketitle

\begin{abstract}
We argue that the velocity dispersions and masses of galactic bulges and spheroids are byproducts of the feedback that regulates rapid black hole growth in protogalaxies. We suggest that the feedback energy liberated by accretion must pass through the accreting material, in an energy-conserving flux close-in and a momentum-conserving flux further out.  If the inflowing gas dominates the gravitational potential outside the Bondi radius, feedback from Eddington-limited accretion drives the density profile of the gas to that of a singular isothermal sphere.  We find that the velocity dispersion associated with the isothermal potential, $\sigma$, increases with time as the black hole mass $M$ grows, in such a way that $M \propto\sigma^4 $.  The coefficient of this proportionality depends on the radius at which the flow switches from energy conserving to momentum conserving, and gives the observed $M-\sigma$ relation if the transition occurs at $\sim 100$ Schwarzschild radii.  We associate this transition with radiative cooling and show that bremsstrahlung, strongly boosted by inverse Compton scattering in a two-temperature ($T_p \gg T_e$) plasma, leads to a transition at the desired radius.  

According to this picture, bulge masses $M_b$ are insensitive to the virial masses of their dark matter haloes, but correlate linearly with black hole mass.  Our analytic model also explains the $M_b-\sigma$ (Faber--Jackson) relation as a relic of black hole accretion.  The model naturally explains why the $M-\sigma$ relation has less scatter than either the $M-M_b$ (Magorrian) or the Faber--Jackson relation. It suggests that the $M-\sigma$ relation could extend down to very low velocity dispersions, and predicts that the relation should not evolve with redshift.  
\end{abstract}
 

\section{Introduction}

Models offering explanations for the ``$M-\sigma$" relation --- the tight correlation between the mass $M$ of the central black hole in a galactic nucleus and the velocity dispersion $\sigma$ of the galaxy's bulge or spheroid (Ferrarese \& Merritt 2000; Gebhardt at al. 2000; Tremaine et al. 2002) ---  fall into two broad categories.  Global models typically appeal to feedback associated with the growth of the black hole.  Kinetic energy (Silk \& Rees 1998; Blandford 1999), radiation energy (Sazonov et al. 2005), bulk momentum (Fabian 1999; King 2003), or radiation pressure (Fabian, Wilman \& Crawford 2002; Murray, Quataert \& Thompson 2005) liberated by accretion reacts back on the infalling material over large distances, stopping both black hole accretion and star formation in the galaxy when the black hole reaches a certain size.  Focusing on the response of the galaxy on scales well outside the black hole's gravitational sphere of influence, these models have paid relatively little attention to the details of the accretion process itself. In contrast, local models relate the growth of the black hole to details of its immediate nuclear environment (e.g., Zhao, Haehnelt \& Rees 2002; Adams et al. 2003; Miralda-Escud\'e \& Kollmeier 2005), but typically have focused on dynamical processes rather than feedback effects.

Recently di Matteo, Springel \& Hernquist (2005) have linked local and global feedback effects, using computer simulations of merging galaxies (see also Springel, di Matteo \& Hernquist 2005).  Although their phenomenological realizations of star formation, radiative cooling in a multiphase medium, and black hole accretion and feedback are extremely crude, they are able to reproduce the $M-\sigma$ relation remarkably well. Interestingly, they find that the $M-\sigma$ correlation is insensitive to the gas fraction in their model galaxies, and that the details of star formation and supernova feedback also have little effect on the black hole mass. These results suggest that the feedback regulating black hole growth operates on local scales --- near the black hole's accretion radius or closer in --- rather than solely on the global scales usually considered.  The insensitivity to gas fraction presumably results because the gas mass is somehow ``maximized" on the scales where the accretion rate is determined; likewise, on these scales black hole feedback is far more important than that due to stars.  
    
In this paper, we present a simple analytic model for the local effects of feedback due to growth of a black hole in a galactic nucleus. When the gas supply is plentiful, the black hole is able to accrete at (or slightly above) the Eddington limit, but feedback modifies the gas flow pattern out to large distances, far beyond the Bondi radius $r_B = GM/\sigma^2$.  If the gas dominates the gravitational potential at $r > r_B$, feedback forces the density distribution to approximate a singular isothermal sphere, $\rho \propto r^{-2}$, with a unique velocity dispersion that satisfies 
$\sigma \propto M^{1/4}$.  We show that the constant of proportionality, which depends on fundamental physical constants and processes occurring close to the black hole, is consistent with the observed $M-\sigma$ relation under plausible assumptions about the gas microphysics.  Remarkably, we find that the $M-\sigma$ relation is satisfied not just for the final black hole mass, but at every stage during black hole growth: $\sigma$ increases along with $M$. 

The plan of the paper is as follows. In \S 2 we discuss the assumptions of the model and derive the main features of the gas flow pattern.  Because of the simple nature of the model --- we invoke turbulent and transsonic energy transport processes using simple scaling laws --- we do not attempt to derive very detailed results. We find that the flow depends on a transition radius between energy-conserving feedback, at small radii, and momentum-conserving feedback at large radii.  We associate this transition with radiative cooling, and in \S~3 we demonstrate that the flow is likely to cool via strongly Comptonized bremsstrahlung at $\sim 100$ Schwarzschild radii, under plausible assumptions about the gas microphysics.  The fact that much of the feedback energy goes into radiation, albeit at large radii, limits to feedback energy flux to roughly the Eddington limit.  In \S 4 we discuss the implications of the model for rapid black hole growth at close to the Eddington limit, arguing that $\sigma$ should evolve with $M$ in such a way as to preserve $M\propto \sigma^4$ at all times.  We consider the growth of the bulge or spheroidal component of the galaxy, as a relic of feedback-regulated accretion, in \S 5.  We argue that the bulge mass (as opposed to the velocity dispersion) is regulated by the potential of the dark matter halo. However, the properties of the bulge depend very weakly on the virial mass of the halo, and the resulting bulge mass $M_b$ turns out to scale roughly linearly with the black hole mass: $M_b \sim 10^3 M$, consistent with observations (Magorrian et al. 1998).  The $M-\sigma$ and $M-M_b$ relations, together, reproduce the Faber--Jackson (1976) relation linking $M_b$ and $\sigma$.  We summarize our main results in \S 6, where we also discuss several predictions and further implications for galaxy formation theory.

\section{Feedback Model}

Suppose an accreting black hole of mass $M$ liberates energy at a rate $L_f = f \dot M c^2 = \ell L_E (M)$, where $L_E (M)$ is the Eddington limit corresponding to mass $M$.  We suppose that the accretion flow is radiatively inefficient close to the black hole; therefore, one might expect the flow to resemble an ADIOS (Adiabatic Inflow-Outflow Solution: Blandford \& Begelman 1999, 2004) and to develop a powerful outflow.  However, we also assume that the incoming gas has sufficiently low angular momentum that it can be regarded as roughly spherically symmetric.  We posit that under such conditions the surrounding gas stifles the wind, i.e., the liberated mechanical energy does not establish a separate channel through which to escape to large distances before interacting with the ambient gas.  Instead, the power that would normally drive the ADIOS wind must pass through the inflowing gas on its way out.  

We envisage two ways for the accreting gas to carry the feedback energy flux without being blown away.  The first can be regarded as a form of convection. Let $L_f \sim 4\pi \rho v_t^3 r^2$, where $v_t$ is a characteristic eddy speed (assumed to be the same as the transport speed) for the energy flux. The characteristic energy transport speed cannot exceed the Keplerian speed, $v_K = c (r/r_g)^{-1/2}$, without blowing apart the inflow.  Adopting this estimate for $v_t$, we find that the density obeys
\begin{equation}   
\rho (r) \sim  {\ell m_p \over r_g \sigma_T} x^{-1/2}, 
\label{rhoin}
\end{equation}
where $m_p$ is the mass of a proton, $\sigma_T$ is the Thomson cross section, $r_g = GM/c^2$ is the gravitational radius of the black hole and $x \equiv r/r_g$. This is analogous to the result obtained by Gruzinov (2002), for non-radiative spherical accretion with phenomenological (transsonic) heat conduction.  It is also analogous to the density law in the CDAF (Convection-Dominated Accretion Flow) model for rotating, radiatively inefficient flows dominated by ion pressure (Narayan, Igumenshchev \& Abramowicz 2000; Quataert \& Gruzinov 2000). In both cases the accreting gas is required to carry a conserved outward energy flux by mechanical means. 

Defining the net inflow speed by $4\pi \rho v_{\rm in} r^2 = \dot M$, we have $v_{\rm in} \sim f^{-1} c x^{-3/2}$.  For $f$ not too small, $v_{\rm in}$ quickly becomes much smaller than $v_t$.  The inflow is very subsonic, nearly stopped by the energy flux it is forced to carry.  From the condition of hydrostatic equilibrium, the gas pressure must obey $p \sim \rho GM/r \propto r^{-3/2}$. We then find that $p/ \rho^{5/3}\propto r^{-2/3}$, i.e., the specific entropy is a strongly decreasing function of $r$.  This means that the inflow is strongly convectively unstable.  The outward heat flux can be regarded as a very inefficient, highly saturated convective flux, making it reasonable to assume that $v_t$ is close to the maximum possible speed, $v_K$. 

The second mode of energy transport is relevant where radiative cooling is important.  Under the conditions we envisage this occurs in the outer parts of the flow, outside a ``cooling radius" $r_c$.  At $r\ga r_c$ much of the convective energy flux is radiated away, but the associated momentum flux, $\Pi_f (r_c) \sim L_f/ v_t(r_c)$, is retained since radiation carries away relatively little momentum.  Because of the rapid cooling, gas pressure alone cannot support the flow against free-fall collapse.  To maintain near-hydrostatic equilibrium, the momentum flux must therefore satisfy
\begin{equation}
{d \Pi_f \over dr} \sim - 4\pi \left( 1 - {L\over L_E}\right) \rho (r) g(r) r^2,
\label{momflux}
\end{equation}
where $g(r)$ is the local gravitational acceleration and $L \sim \ell L_E$ is the radiative luminosity liberated at $r \ga r_c$.  We visualize $\Pi_f (r)$ as an internal flux of momentum, carried by predominently radial turbulent motions, rather than as an outflowing wind.

The factor $1 - L/L_E$ at in eq.~(\ref{momflux}) implies that the feedback energy flux cannot exceed the Eddington limit, even though the flow is assumed to be radiatively inefficient close to the black hole.  Since we are interested in rapid accretion in the gas-rich environment of a protogalaxy, we will assume that the flow is Eddington-limited and that $\ell \approx 1$. Since $\Pi_f (r_c) \sim 4\pi \rho(r_c) g(r_c) r_c^3$, according to eq.~(\ref{momflux}) the momentum flux should be approximately conserved at $r> r_c$.  

We define the Bondi radius $r_B$ as the place where the gravitational acceleration due to the black hole equals $g_{\rm gal}(r)$ due to the stars, gas, and dark matter in the galaxy (but not including the black hole).  At $r > r_B$ the Eddington limit associated with the enclosed mass exceeds $L_E (M)$, hence we can set $1 - L/L_E \approx 1$ in eq.~(\ref{momflux}).  Integrating eq.~(\ref{momflux}) from $r_B$ to some larger radius $r$, we obtain 
\begin{equation}
\Pi_f (r) \sim \Pi_f (r_c) - 4\pi \int^r_{r_B} \rho g r^2 dr   ,
\label{momint}
\end{equation}
with $g \approx g_{\rm gal}(r)$.  We identify three qualitatively distinct cases. If $\rho g r^3$ increases as a power of $r$, the momentum flux will be quickly depleted and the feedback will fail.  Matter will slump toward the center until $\rho g r^3$ is roughly independent of $r$. On the other hand, if $\rho $ decreases so rapidly that $d \Pi_f/dr \approx 0$, there will be enough momentum flux to drive a strong outflow at $r > r_B$.  Once the local mass supply is depleted the accretion rate, and therefore the feedback, will decline, allowing matter to fall toward the black hole and increase $\rho$.   We assume that the third case, intermediate between the first two, is the one that is chosen by nature.  We conjecture that $\rho (r)$ adjusts itself so that $\rho(r) g(r) r^3 \sim \Pi_f(r_c) \sim$ constant (to within a factor $\sim \ln r$).  This self-regulation is the crux of our feedback argument; its validity will have to be checked by more detailed means --- presumably through numerical simulations.  In \S~4 we will explore its ramifications, but first we discuss the value of the cooling radius, $r_c$.

\section{Cooling Radius}

As we shall see in \S 4, the coefficient of the predicted $M-\sigma$ relation depends on the radius at which the feedback makes a transition from energy-conserving to momentum-conserving, and that our predictions match the observations for $x_c \sim 100$.  In this section we show that such values of $x_c$ are plausible, provided that two conditions are met: 1) the electrons and ions are coupled thermally by Coulomb scattering and are permitted to have different temperatures; and 2) the dominant cooling mechanism is thermal bremsstrahlung, strongly amplified by inverse Compton scattering.  
 
We adopt the density profile given by eq.~(\ref{rhoin}) with $\ell = 1$ and assume that the protons track the virial temperature, $k T_p/m_p c^2 \approx x^{-1}$. Defining a dimensionless electron temperature $\theta_e \equiv kT_e/m_e c^2$ and dividing volume heating and cooling rates by $m_p c^3/(r_g^2\sigma_T)$, we obtain the dimensionless Coulomb energy transfer rate from protons to electrons, 
\begin{equation}
{\cal C}_{\rm pe} \sim 0.013 \left( x^{-1} - {m_e\over m_p} \theta_e \right)\theta_e^{-3/2} x^{-1}
\label{cipe}
\end{equation}
(Stepney \& Guilbert 1983, where we have assumed a Coulomb logarithm $\lambda \sim 20$).  Our justification for assuming strong Comptonization comes from the fact that the flow at $r\la r_c$ is very optically thick to electron scattering, with a characteristic optical depth $\tau (x) \sim x^{1/2}$, and Coulomb coupling is likely strong enough to maintain an electron temperature $\theta_e \ga 0.1$. Thus, the Compton ``$y-$parameter", $y \approx 4\theta_e(1 + 4\theta_e) \tau^2 $ (Rybicki \& Lightman 1979), is large and Comptonization is highly saturated, with low-frequency photons boosted into a Wien peak at the inverse Compton temperature (which equals the electron temperature, under these conditions).  The dimensionless bremsstrahlung cooling rate is given by 
\begin{equation}
{\cal C}_{\rm cbr} \sim 2.9 \times 10^{-6} {\cal A} \ \theta_e^{1/2} x^{-1},
\label{cibr}
\end{equation}
where $\cal A$ is the amplification factor due to Comptonization of the bremsstrahlung, discussed at length by Rybicki \& Lightman (1978: see their eq.~[7.74] and preceding discussion).  The cooling radius is defined by the condition that the volume-integrated bremsstrahlung emissivity equal $L_f$; this is approximately equivalent to the local condition ${\cal C}_{\rm cbr} (x_c) = 3 x_c^{-3}$.  Simultaneously demanding ${\cal C}_{\rm pe} = {\cal C}_{\rm cbr}$, we can solve for $x_c$ and the electron temperature $\theta_e (x_c)$.  The result is weakly dependent on $M$ --- applying the prescription in Rybicki \& Lightman (1978) we obtain: $x_c = 74, \ 68, \ 63, \ 58$; $\theta_e(x_c) = 0.46, \ 0.44, \ 0.42, \ 0.40$ for $M=10^4, \ 10^6, \ 10^8, \ 10^{10} \ M_\odot$, respectively. The Compton amplification factor ranges from 275 to 487, over the same range of masses.    

At $r<r_c$, the electrons are nearly isothermal at $\theta_e (x_c)$, while at larger radii the continued importance of Comptonization and bremsstrahlung emission (which creates a large number of photons at low energies, even if the total emissivity is declining) is likely to lead to a decreasing Compton temperature for some range of $r>r_c$.  The optical depths and densities in the flow are so large that expulsion of the gas at large distances by X-ray pre-heating is unlikely (Sazonov, Ostriker \& Sunyaev 2004; Sazonov et al. 2005).  For example, the ionization parameter $\xi = L/nr^2$ is $\la 10$ at $r > r_B$. 
 
Cooling by Comptonized thermal cyclotron emission is unlikely to be as important as bremsstrahlung under the assumptions of our model. To estimate the contribution from cyclotron losses, we assume that the magnetic energy density is $\sim 10\%$ of the ion pressure. The emission at low harmonics is heavily self-absorbed, and we find that the dominant source of seed photons for Comptonization comes from emission at harmonics $n_{\rm cyc} \la 75$ (Bekefi 1966; Takahara \& Tsuruta 1982).  The Comptonized emissivity scales with radius roughly $\propto x^{-3}$, implying that cyclotron emission either cools the flow close to the black hole, or not at all.  Seed photon emission from high harmonics increases the Comptonized emissivity by a factor $\sim n_{\rm cyc}^2$; nevertheless, we find that under optimistic conditions, cooling by Comptonized cyclotron emission fails by a factor of a few.  We note that synchrotron radiation and Compton scattering by highly relativistic electrons may also contribute to radiative losses, but since they deplete only the relativistic (nonthermal) electron population they are unlikely to be decisive in cooling the flow.
 
Therefore, we conclude that it is plausible that Eddington-limited flows dominated by mechanical feedback can remain radiatively inefficient out to $x \sim 60$.  In the remainder of this paper we will treat $x_c$ as a parameter with a fiducial value of 100, allowing readers to gauge the sensitivity of our predictions to details of the gas microphysics.

\section{Rapid Black Hole Growth and the $M-\sigma$ Relation}

Now let us consider the rapid growth of a black hole during the epoch of galaxy formation. We assume that there is so much gas falling into the central regions of the protogalaxy that the gas mass dominates locally over that of both the dark matter and the stars that have formed so far.  At $r > r_B$ we have 
\begin{equation}
g_{\rm gal} (r) \sim 4\pi G  \rho r ,
\label{hydrorb}
\end{equation}
where the factor $4\pi$ anticipates the radial dependence of $ \rho$.  The only thing that can limit the infall of gas is the feedback from the black hole, but at the same time increasing the gas density at $r_B$ will increase the black hole accretion rate.  As noted in \S~2, we assume that the total feedback energy flux reaches a plateau at roughly the Eddington limit.  This limiting feedback applies even if the accretion rate is much larger than the nominal limiting value of $\sim 10 L_E/c^2$ (where we assume an accretion efficiency of $\sim 0.1$): in this case, most of the liberated binding energy would be advected into the black hole (Begelman 1979) and $f$ would be $\ll 1$.   Substituting for $\rho(r) g_{\rm gal}(r) r^3 \sim \Pi_f(r_c)$ from above, we obtain  
\begin{equation}
\rho \sim \left({M m_p\over 4\pi \sigma_T}\right)^{1/2}  x_c^{1/4} r^{-2} .
\label{rhobar}
\end{equation}
Thus, feedback due to Eddington-limited growth of the black hole causes the inflowing gas at $r > r_B$ to distribute itself as a singular isothermal sphere. Note also that $\rho$ does not depend on the thermal state of the accreting medium, i.e., on the details of heating and cooling processes at $r > r_B$.

The velocity dispersion associated with the gas distribution is given by $\sigma^2 \sim 2\pi G \rho r^2 \propto M^{1/2}$. Thus, $M \propto \sigma^4$ at all stages during the Eddington-limited growth of the black hole.  Quantitatively, we can write the proportionality in the forms
\begin{eqnarray}
M \sim m_p \left({\sigma_T c^4 \over \pi G^2 m_p^2 }\right) x_c^{-1/2} \left(\sigma\over c \right)^4 \nonumber \qquad \qquad\qquad & & \\ 
\qquad  \qquad \qquad \sim 2.4 \times 10^8 \left({x_c \over 100} \right)^{-1/2} \sigma_{200}^4 \ M_\odot, 
\label{msigma}
\end{eqnarray}
where $\sigma_{200} = \sigma/200$ km s$^{-1}$.  We note that this relationship is similar but not identical to the expressions derived by King (2003) and by Fabian (1999).  This similarity should not be a surprise, since the feedback effect in all three models relies on momentum balance.  There are two fundamental differences between our model and the earlier ones.  First, in our model the black hole mass determines the velocity dispersion of the bulge, rather than determining the bulge mass.  As we will see below, the bulge mass is determined by the potential of the dark matter halo, once the bulge velocity dispersion is fixed by black hole feedback.  Second, black hole and bulge growth in our model do not stop because the infalling gas is somehow ``blown away" by the feedback.  A given dark matter halo could, in principle, host black holes and bulges with a wide range of masses and velocity dispersions (although, in practice, the supply of gas available to build both bulge and black hole is probably correlated with the mass of the halo).  What our model implies is that $M$, $\sigma$, and (as we shall see) the bulge mass are all correlated, whatever their final values.  Considering the crude approximations used in our analysis, the theoretical correlation (\ref{msigma}) is remarkably close to the observed $M-\sigma$ relation (Tremaine et al. 2002), provided that the transition radius $r_c$ is not too different from the value estimated in \S 3 based on bremsstrahlung cooling.  The symbolic expression is intended to display the dependence of the $M-\sigma$ relation on fundamental parameters.  In particular, the first quantity in parentheses is roughly the square of the ratio of the classical electron radius, $r_0 = e^2/m_e c^2 \sim \sigma_T^{1/2}$, to the gravitational radius of a proton, $Gm_p/c^2$.   

As the gas supply is depleted, for whatever reason --- star formation, mass loss from the outer parts of the system, and/or simply the limited gas content of the halo --- the accretion rate drops below the Eddington limit and the black hole mass levels off.  The final mass of the black hole determines the final value of $\sigma$ associated with the gas distribution and the stellar spheroid that eventually forms from it.  Thus, our simple feedback mechanism explains the observed $M-\sigma$ relation as a relic of the co-evolution of $M$ and $\sigma$.  

\section{Bulge Mass and the Faber-Jackson Relation}     

Although our feedback model relates $\sigma$ to $M$, it does not directly determine the mass of the bulge that grows in tandem with the black hole.  At radii where gas dominates the gravitational potential, the enclosed bulge mass increases as $M_b(<r) \propto  \rho r^3 \sim M (r/r_B)$. Were it not for the existence of a dark matter halo, this behavior could extend out to very large radii (limited mainly by the availability of gas) and the bulge could be many orders of magnitude more massive than the hole.  However, if the dark matter density distribution is shallower than that of the gas, as it is believed to be in the central regions of galaxies (Navarro, Frenk \& White 1997, hereafter NFW), it will eventually dominate the potential.  Where the inner NFW density distribution, $\rho_{\rm NFW}\propto r^{-1}$, dominates the potential, we have $ \rho \propto r^{-3}$ from the momentum conservation condition.  The enclosed bulge mass thus goes from increasing linearly with $r$ to increasing logarithmically.  The radius where $\rho_{\rm NFW} =  \rho$ thus determines the mass of the bulge.

To evaluate the bulge mass predicted by our model, we assume the ``concordance" $\Lambda$CDM model with (at $z=0$) $\Omega_m = 0.3$, $\Omega_\Lambda = 0.7$, and $h=0.65$.  The inner NFW density profile is then given by  
\begin{equation}
\rho_{\rm NFW} (r) \approx 9\times 10^{-24} M_{\rm v,12}^{0.07} \  r_{\rm kpc}^{-1} \ [\xi(z)]^{2/3} \ {\rm g \ cm}^{-3} 
\label{rhoNFW}  
\end{equation}
(Komatsu \& Seljak 2001), where $M_{\rm vir} = 10^{12} M_{\rm v,12} M_\odot$ is the virial mass of the halo, $\xi(z) = [\Omega_m / \Omega_m(z)] (18\pi^2 + 82x- 39 x^2)/100$ with $x \equiv \Omega_m(z) - 1$ (Bryan \& Norman 1998), and we have used $c \approx 12.8 M_{\rm v,12}^{-0.13} h^{-0.13} (1+z)^{-1}$ (Bullock et al. 2001) for the concentration parameter.

Equating $\rho_{\rm NFW}$ to $ \rho$ we obtain the bulge radius,
\begin{equation}
r_b \sim 7.4 \left({x_c \over 100} \right)^{1/4} M_8^{1/2} \ M_{\rm v,12}^{-0.07} \ [\xi(z)]^{-2/3}  \ {\rm kpc}, 
\label{rbulge}  
\end{equation}
where $M_8 = M/10^8 M_\odot$, and the ratio of bulge mass to black hole mass,
\begin{eqnarray}
{M_b \over M }  \sim  {r_b \over r_B}\sim  550  \left({x_c \over 100}\right)^{1/2}  M_{\rm v,12}^{-0.07} \ [\xi(z)]^{-2/3}, 
\label{Mbulge}  
\end{eqnarray}
where we have used eq.~(\ref{msigma}).  Thus, we find an approximate proportionality between the mass of the black hole and the mass of the bulge, roughly consistent with the Magorrian et al. (1998) relation.  The virial mass of the dark matter halo enters only to the -0.07 power.  The $M-\sigma$ relation then gives $M_b \propto \sigma^4$, recovering the Faber--Jackson (1976) relation for spheroids. 

Although the $M-\sigma$ relation is independent of redshift, the ratio between $M$ and the bulge mass $M_b$ can exhibit weak $z-$dependence due to the factor $[\xi(z)]^{-2/3}$. For the concordance model with a cosmological constant, $\xi(z)$ decreases from $\approx 1$ at $z = 0$ to $18 \pi^2 \Omega_m/ 100 \approx 0.53$ as $z \rightarrow \infty$.  For $z < 0.5$, $\xi(z) \approx (1 + z)^{-0.8}$.  The general trend is for $M_b/M$ to increase slightly with  redshift, but by less than a factor of 2.  For $\Lambda$CDM, $M_b/M$ is larger by about 30\% at $z = 0.5$, 40\% at $z=1$, and 50\% at $z \geq 5$. This modest trend will be further weakened by the fact that black holes observed at a given $z$ formed over a range of redshifts.

\section {Discussion and Conclusions}
 
We have described a simple feedback model relating the Eddington-limited growth of supermassive black holes in young galaxies to the properties of the inflowing gas. The model requires that the feedback fluxes of energy and momentum pass through the material surrounding the black hole, on all scales except perhaps in the immediate vicinity of the event horizon.  This implies that the incoming gas not hit a centrifugal barrier until approaching the black hole.  Although the sub-Keplerian angular momentum of the flow probably reflects boundary conditions in the outer parts, we speculate that the turbulent fluxes of energy and momentum sustain quasi-spherical flow at smaller radii by transporting angular momentum outward.  

The model predicts that, once the inflowing gas dominates the gravitational potential in the galaxy's core, feedback forces the gas to develop the density profile of a singular isothermal sphere, with a velocity dispersion that depends on the instantaneous mass of the black hole.  In contrast to most other feedback models, which assume that the black hole grows to a certain limiting mass in a fixed galactic potential, our model implies that the velocity dispersion associated with the bulge, $\sigma$, increases as $M$ increases, in such a way that $M \propto \sigma^4$ at all times.    

Our model reproduces three observed correlations: the $M-\sigma$ relation, the Magorrian ($M-M_b$) relation, and the Faber--Jackson ($M_b-\sigma$) relation.  If it is correct, the model implies that spheroids form as the result of self-regulated accretion onto supermassive black holes and grow in tandem with them.  Of the three correlations, the $M-\sigma$ relation is the most direct, and therefore it should be no surprise that it seems to be the tightest observationally.  Both the Magorrian and Faber--Jackson correlations also involve the properties of the dark matter halo, and therefore should exhibit more scatter.  

Our feedback model makes some clear predictions and has a number of interesting consequences for galaxy formation.  Since the coefficient of proportionality between $M$ and $\sigma^4$ depends firstly on the maximum feedback luminosity (which we assume is close to the Eddington limit) and secondly on gas microphysics close to the black hole (where electron-ion coupling via Coulomb scattering and Comptonized bremsstrahlung determine the cooling radius), there should be no limit to the range of velocity dispersions over which the $M-\sigma$ relation can apply. The only requirements are that the growth of the black hole be dominated by Eddington-limited accretion and that the feedback operate according to our scheme.  (These conditions could be violated, for example, if the inflowing gas were rapidly rotating at large radii or if mergers of smaller black holes had played a large role in black hole growth.)  In particular, the $M \propto \sigma^4$ scaling could extend to extremely low velocity dispersions, where cooling rates at $r_B$ differ dramatically from that of bremsstrahlung. 

The model predicts that there should be no dependence of the $M-\sigma$ relation on redshift. However, the ratio between the bulge mass $M_b$ and $M$ could show a weak trend with $z$, depending on cosmological parameters.  For the concordance $\Lambda$CDM model with a cosmological constant the trend should be toward higher ratios $M_b/M$ with increasing $z$.  Black holes and bulges forming at $z=0.5$ could have $M_b/M\sim 30$\% higher than black holes and bulges forming at $z\sim 0$; $M_b/M$ could be as much as 50\% higher for black holes forming at $z>5$, but the amplitude of the effect would be significantly reduced if the black holes observed at $z_{\rm obs}$ formed over a range of redshifts much larger than $z_{\rm obs}$.
 
Although the ratio $M/M_b$ is well-determined by the model, the final mass of the black hole, and of the bulge, could be sensitive to processes that trigger the rapid inflow of gas into the center of the halo. These may depend on the details of galaxy mergers or tidal encounters, and lead to a large scatter in, e.g., ratios of black hole or bulge mass to halo mass or circular velocity.  This could explain why black hole masses appear to correlate with bulge properties but not with the properties of disks (Kormendy 2001).  However, it is likely that the supply of gas available for growing the bulge and black hole is to some extent correlated with the mass of the halo. A linear correlation appears to be necessary in order to explain the entropy excess of galaxy groups and clusters in terms of the energy injected by a central black hole (Roychowdhury et al. 2004).

The model also predicts that rapidly accreting black holes should be heavily obscured, with a characteristic column density $N \sim 1.5 \times 10^{25} (x_c / 100)^{1/2}$ cm$^{-2}$, independent of $M$. Black holes accreting at close to the Eddington limit are therefore expected to be Compton-thick, as previously discussed by Fabian (1999; see also Fabian et al. 2002) and Hopkins et al. (2005).   
 
The feedback mechanism we have described should also be relevant under certain conditions when the gas outside $r_B$ is not self-gravitating and/or the black hole is accreting at less than the Eddington limit. Possible consequences for the fueling of active galactic nuclei will be discussed elsewhere.

Feedback mechanisms like the one we have described, in which the feedback energy must pass through the inflowing matter in a (presumably) highly turbulent fashion, present a significant challenge to numerical modelers.  From the perspective of state-of-the-art models like that recently described by di Matteo et al. (2005: see also Springel et al. 2005), our feedback process represents ``subgrid" physics.  However, the basic physical elements of feedback, radiative cooling, and conserved momentum and/or energy fluxes are present in both treatments, and it is not impossible that some of the effects we describe are implicit in their simulations.  

\section*{Acknowledgments} 
We thank Mateusz Ruszkowski and Martin Rees for stimulating discussions and useful suggestions. This work was supported in part by National Science Foundation grant AST-0307502. BBN thanks the Fellows of JILA for their hospitality.

\end{document}